\documentclass[aps,prb,twocolumn,superscriptaddress,showpacs]{revtex4}

\usepackage{graphicx}
\usepackage{dcolumn}
\usepackage{bm}
\usepackage{amssymb}
\usepackage{amsmath}

\usepackage{color}
\newcommand{\lco}{LaCoO$_3$}
\newcommand{\lsco}{La$_{0.998}$Sr$_{0.002}$CoO$_3$}
\newcommand{\xlsco}{La$_{1-x}$Sr$_{x}$CoO$_3$}

\newcommand{\etal}{\textit{et al.}}

\begin{document}

\title{Effect of carrier doping on the formation and collapse of magnetic polarons in lightly hole-doped \xlsco}



\author{A.~Podlesnyak}
\thanks{Corresponding author. Electronic address: podlesnyakaa@ornl.gov}
\affiliation{Neutron Scattering Science Division, Oak Ridge National Laboratory, Oak Ridge, Tennessee 37831, USA}
\author{G.~Ehlers}
\affiliation{Neutron Scattering Science Division, Oak Ridge National Laboratory, Oak Ridge, Tennessee 37831, USA}
\author{M.~Frontzek}
\affiliation{Neutron Scattering Science Division, Oak Ridge National Laboratory, Oak Ridge, Tennessee 37831, USA}
\author{A. S. Sefat}
\affiliation{Materials Science and Technology Division, Oak Ridge National Laboratory,  Oak Ridge, Tennessee 37831, USA}
\author{A.~Furrer}
\affiliation{Laboratory for Neutron Scattering, Paul Scherrer Institut, CH-5232 Villigen PSI, Switzerland}
\author{Th.~Str{\"a}ssle}
\affiliation{Laboratory for Neutron Scattering, Paul Scherrer Institut, CH-5232 Villigen PSI, Switzerland}
\author{E.~Pomjakushina}
\affiliation{Laboratory for Developments and Methods, Paul Scherrer Institut, CH-5232 Villigen PSI, Switzerland}
\author{K.~Conder}
\affiliation{Laboratory for Developments and Methods, Paul Scherrer Institut, CH-5232 Villigen PSI, Switzerland}
\author{F.~Demmel}
\affiliation{ISIS Facility, Rutherford Appleton Laboratory, Didcot OX11 0QX, United Kingdom}
\author{D.~I.~Khomskii}
\affiliation{II~Physikalisches Institut, Universit{\"a}t zu
K{\"o}ln, Z{\"u}lpicher Stra{\ss}e 77, 50937 K{\"o}ln, Germany}

\date{\today}

\begin{abstract}
We investigate the doping dependence of the nanoscale electronic and magnetic inhomogeneities in the hole-doping range ${0.002}\leqslant{x}\leqslant{0.1}$ of cobalt based perovskites, \xlsco.
Using single crystal inelastic neutron scattering and magnetization measurements we show that the lightly doped system exhibits magneto-electronic phase separation in form of spin-state polarons.
Higher hole doping leads to a decay of spin-state polarons in favor of larger-scale magnetic clusters, due to competing ferromagnetic correlations of Co$^{3+}$ ions which are formed by neighboring polarons.
The present data give evidence for two regimes of magneto-electronic phase separation in this system: (i) ${x}\lesssim{0.05}$, dominated by ferromagnetic intrapolaron interactions, and (ii) ${x}\gtrsim{0.05}$, dominated by Co$^{3+}$-Co$^{3+}$ intracluster interactions.
Our conclusions are in good agreement with a recently proposed model of the phase separation in cobalt perovskites [He \etal, Europhys. Lett. \textbf{87}, 27006 (2009)].
\end{abstract}


\maketitle

\section{INTRODUCTION}

It is generally accepted that the development of a magnetoelectronic phase separation (MEPS) in hole-doped perovskite cobaltites \xlsco\ plays a crucial role for their magnetic and transport properties.
As found by various experimental techniques, such as nuclear magnetic resonance,\cite{Kuhns,Hoch,Podlesnyak3} small-angle scattering,\cite{Wu}
diffraction,\cite{Caciuffo}
inelastic neutron scattering (INS),\cite{Phelan,Phelan1,Podlesnyak3}
transmission electron microscopy,\cite{Caciuffo}
extended x-ray absorption fine structure (EXAFS) measurements,\cite{Sundaram}
magnetometry\cite{Giblin1,Giblin2}
and heat capacity measurements,\cite{He2,He3} the phase separation leads to the formation of ferromagnetic (FM) clusters in a nonferromagnetic matrix upon carrier injections.
Recent theoretical efforts,\cite{Dagotto2,Kugel,Sboychakov,Suzuki} mainly based on the early "ferrons'' ideas of Nagaev,\cite{Nagaev1,Nagaev2,Nagaev3} contributed to considerable progress in the understanding of the origin of electronic phase separation in a wide doping range.
The consensus is that the low temperature phase diagram starts from a nonmagnetic state at $x=0$, and upon doping includes two large regions of spin-cluster glass (SG) and ferromagnetic (FM) states with a metal-insulator transition (MIT)
at ${x}\thickapprox{0.18-0.22}$.\cite{Wu,Phelan1,Itoh94,G2,He}
The nonmagnetic ground state of the parent compound, \lco, corresponds to a low-spin (LS) state of Co$^{3+}$ ions ($t^6_{2g}e^0_g$, $S=0$).\cite{eae1}
Due to subtle balance between the intra-atomic (Hund's) exchange interaction $E_{\textsc{h}}$ and the crystal-field splitting $\Delta_{cf}$, the first excited state is located at about 10~meV (0.5-0.7~\% of $\Delta_{cf}$)\cite{eae} above the LS state.\cite{Podlesnyak1,Haverkort1,He1}
The closeness in energy of these states makes \lco\ a well-known spin-state transition model system.
The SG state is characterized by a hole-poor nonferromagnetic matrix with embedded hole-rich FM droplets.\cite{Wu}
The nonferromagnetic spin correlations in the matrix,\cite{Phelan2} FM intracluster correlations\cite{Caciuffo} and magnetic interactions between the matrix and FM clusters\cite{Giblin1,Phelan} give rise to the inhomogeneous magnetic nature of \xlsco.

In a recent comprehensive study He \etal\ suggested that MEPS occurs only in a well-defined doping range, ${0.04}\leqslant{x}\leqslant{0.22}$.\cite{He2,He3}
They showed that the phase separation is controlled solely by the site occupation randomness introduced by the doping, and is not electronically driven.\cite{He3}
At increasing $x$ the clusters eventually reach the percolation limit leading first to short-range magnetic order at ${x}\sim{0.18}$, and then long-range FM order at ${x}\sim{0.22}$.\cite{Wu,Phelan1}
At $x>0.22$ the system becomes a ferromagnetic metal, although FM and non-FM clusters coexist well above the MIT in a composition range often characterized as cluster-glass (CG) region.\cite{Itoh94,Wu1,Kuhns}
Results obtained by polarized neutron inelastic scattering in ferromagnetic
La$_{0.82}$Sr$_{0.18}$CoO$_3$ suggest that low-energy spin excitations can be described in terms of a simple localized Heisenberg ferromagnet.\cite{Ewings}

Much less is known about the development of the MEPS around the low limit
of ${x}\thickapprox{0.04}$.
Different experimental techniques proved that the system is phase separated below this limit as well.\cite{Podlesnyak3,Yam1,Giblin1,Smith}
Recently, we elucidated the mechanism of how already the light hole
doping ${x}\sim{0.002}$ dramatically affects magnetic properties of \lco,\cite{Podlesnyak3} an effect first discovered by  Yamaguchi \etal~\cite{Yam1}
Our analysis revealed that the charges introduced by substitution of Sr$^{2+}$ for La$^{3+}$ do not remain localized at the Co$^{4+}$ sites.
Instead, each hole is extended over the neighboring Co$^{3+}$ ions, transforming them to a higher spin state and thereby forming a magnetic spin-state polaron.
Important questions remain: How do the polarons behave with increasing Sr content across the low limit ${x}\thickapprox{0.04}$ proposed for the "true'' magnetoelectronic phase separation border? What is the characteristic distinction between these, both magnetically inhomogeneous, states?

In this work, combining single crystal INS with magnetization data of \xlsco, we present a detailed study of the doping dependence of the magneto-electronic phase separation through the critical limit ${x}\sim{0.04}$.
We give unambiguous evidence that at low doping, ${x}\lesssim{0.04}$, the nanoscale MEPS is stabilized in the form of heptamer FM spin-state polarons in a non-FM matrix.
We also show that further hole doping above the critical limit ${x}\sim{0.04}$ leads to a decay of spin-state polarons mainly due to FM exchange of neighboring Co$^{3+}$ ions of different polarons at the expense of AFM interpolaron interactions.
In turn, this results in the appearance of hole-rich FM clusters in significant size and density, stabilizing the CG region.

\section{EXPERIMENTAL}
\subsection{Sample preparation}

Starting powders for single crystal growth of \xlsco, $x=0, 0.002, 0.005, 0.01, 0.02, 0.05, 0.1$ were synthesized by a solid state reaction using La$_2$O$_3$, SrCO$_3$ and Co$_3$O$_4$ of a minimum purity of 99.99\%.
Stoichiometric amounts of the oxides and carbonates were ground thoroughly and fired at temperatures from 850 to 1200$^{\circ}$C several times.
The complementary powders of La$_{1-x}Me_{x}$CoO$_3$, $Me=$~Ca, Y, were prepared as well using appropriate carbonate and oxide (CaCO$_3$, Y$_2$O$_3$).
About $\sim50$~g of each polycrystalline sample was prepared, and therefore the absolute mass of the dopant could be weighed with sufficient accuracy.
The phase purity of the synthesized compounds was verified by means of powder x-ray diffraction.
Single crystals were grown using an Optical Floating Zone Furnace.

Particular attention was paid to the oxygen stoichiometry of the samples and the homogeneity of the Sr doping along the length of the as-grown  crystals.
The oxygen content of polycrystalline and single crystal samples was determined by thermogravimetric hydrogen reduction.\cite{Conder}
The oxygen nonstoichiometry, which can produce effects similar to Sr doping, was found to be less than 0.01.
The small concentration of the doping element made it difficult to control the distribution of strontium along the grown crystals with laboratory techniques such as x-ray diffraction or energy dispersive x-ray analysis alone.
It is known that the low temperature magnetic susceptibility for lightly doped samples is remarkably increased with such light doping.\cite{Yam1}
Therefore, we compared the temperature dependent magnetization of small crystal pieces taken from different places of the as-grown crystals.
The results of the magnetization measurements obtained for all the crystal pieces and also for starting powder were identical (within each strontium concentration) and consistent with previously published data,
where available.\cite{Yam1}
This proved that (i) the strontium distribution was homogeneous throughout the sample volume and (ii) the actual dopant concentration was close to the nominal value.

\subsection{Instrumentation}

INS experiments on single crystal samples of \xlsco\ were performed at the Cold Neutron Chopper Spectrometer (CNCS) at the Spallation Neutron Source in Oak Ridge,\cite{CNCS} at the backscattering spectrometer IRIS at the ISIS neutron scattering facility, and at the time-of-flight spectrometer FOCUS at the Swiss spallation neutron source SINQ at PSI.\cite{FOCUS}
For the CNCS measurements the single crystal ($x=0.01$) was mounted in the (H,H,L) scattering plane (throughout the paper we use the pseudocubic notation with the scattering wave vector $\mathbf{Q}$, given in reciprocal lattice units (r.l.u.)).
The measurements were done at temperatures from 1.5 to 40~K with an incident neutron energy $E_i=3.7$~meV.
At this energy the instrumental elastic energy resolution, full width at half maximum, was 70~$\mu$eV.
The data at FOCUS ($x=0.002, 0.005, 0.01$) were collected in the (H,K,0) scattering plane using an incident neutron energy $E_i=3.55$~meV, giving an elastic resolution of 150~$\mu$eV.
The IRIS experiment (samples with Sr concentration $x=0.05$ and $0.1$ in the (H,H,L) scattering plane) used the pyrolitic graphite (004) reflection to select a fixed final energy $E_f=7.38$~meV resulting in a
resolution of $\sim{55}$~$\mu$eV at the elastic position.
The data were corrected for detector efficiency using a vanadium standard.
The program MSLICE ported in the DAVE software package was used for data visualization and analysis.\cite{Mslice}
Magnetization measurements were performed using a SQUID MPMS magnetometer.

\section{RESULTS AND DISCUSSION}
\subsection{Inelastic Neutron Scattering}
\subsubsection{\xlsco, ${x}\leqslant{0.01}$; the case of weakly interacting spin-state polarons}

No magnetic excitations have been found for temperatures $T<30$~K in the parent compound \lco.\cite{Podlesnyak1}
An inelastic peak at $\delta E=0.6$~meV was found at intermediate temperatures starting from $T \sim 30$~K.
This excitation is due to a thermally excited HS magnetic state of Co$^{3+}$ ions in the non-disturbed \lco\ matrix, as was discussed in our previous
work.\cite{Podlesnyak1}
Earlier INS experiments on lightly hole-doped polycrystalline \xlsco\
(${x}\sim{0.002}$) provided evidence for the existence of octahedrally shaped polarons which consist of a central Co$^{4+}$ ion in LS state configuration, $S_1=1/2$, surrounded by six Co$^{3+}$ ions along the three cube axes in intermediate spin state with spin $S_2=S_3= \ldots =S_7=1$.\cite{Podlesnyak3}
This conclusion was largely based on the $Q$ dependence of the magnetic peak intensity observed for a polycrystalline sample of \lsco\ at energy transfer $\delta=0.75$~meV and also on the Zeeman splitting of this peak in magnetic field, see Fig.~1 in Ref.~\onlinecite{Podlesnyak3}.
This peak in the INS spectrum corresponds to the transition between the ground state levels of the Co heptamer split by a weak trigonal crystal field.
The details of the nature of this magnetic excitation will be discussed in the  Appendix.
The neutron cross section for polycrystalline samples can be written down as a superposition of damped sine functions:
\begin{equation}
\frac{d^2\sigma}{d\Omega d\omega} \quad\propto\quad F^2(Q)
\sum_{j<j{^\prime}=1}^n \Bigg(1+2 \frac{\sin(Q|R_j-R_{j\prime}|)} {Q|R_j-R_{j\prime}|} \Bigg), \label{average}
\end{equation}
where $F(Q)$ is the magnetic form factor, $Q$ the modulus of the scattering vector, and $R_j$ denotes the position vector of the $j$th Co ion.
Although the data displayed in Ref.~\onlinecite{Podlesnyak3} are best described by a Co heptamer (n=7), other types of magnetic clusters cannot be excluded unambiguously.
In particular, the $Q$ dependencies of the magnetic intensities for the heptamer and octamer (n=8) clusters are rather similar.
The values of the saturated magnetic moments, $13\mu_{\textsc{b}}$ for n=7 and $15\mu_{\textsc{b}}$ for n=8, are also too close to be distinguished by magnetization measurements.
However, a clear-cut discrimination is possible by studying single crystals. In this case the neutron cross section has the form
\begin{equation}
\frac{d^2\sigma}{d\Omega d\omega} \quad\propto\quad F^2(\mathbf{Q})
\sum_{j<j{^\prime}=1}^n \Bigg(1+2 \cos{\mathbf{Q}(\mathbf{R}_j-\mathbf{R}_{j\prime})}\Bigg). \label{single}
\end{equation}

\begin{figure}[tb!]
\begin{center}
\includegraphics[width=0.8 \columnwidth]{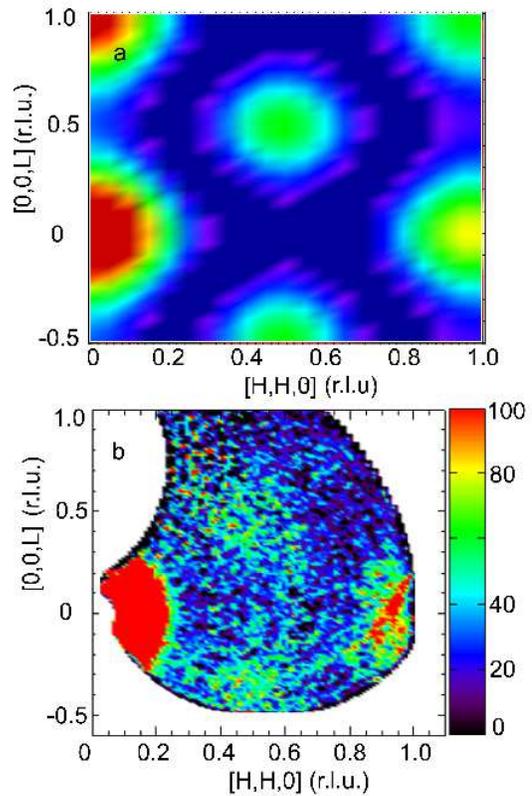}
\caption {(Color online) Constant energy map of the (a) calculated intensity for the case of octahedral 7-site magnetic cluster and (b) measured on CNCS inelastic neutron scattering intensity of the corresponding peak at $\delta E=0.75$~meV in the (H, H, L) scattering plane of reciprocal space obtained from La$_{0.99}$Sr$_{0.01}$CoO$_3$ at $T=2$~K.
The intensities are in arbitrary units. \label{INS_map}}
\end{center}
\end{figure}

\begin{figure}[tb!]
\begin{center}
\includegraphics[width=0.95 \columnwidth]{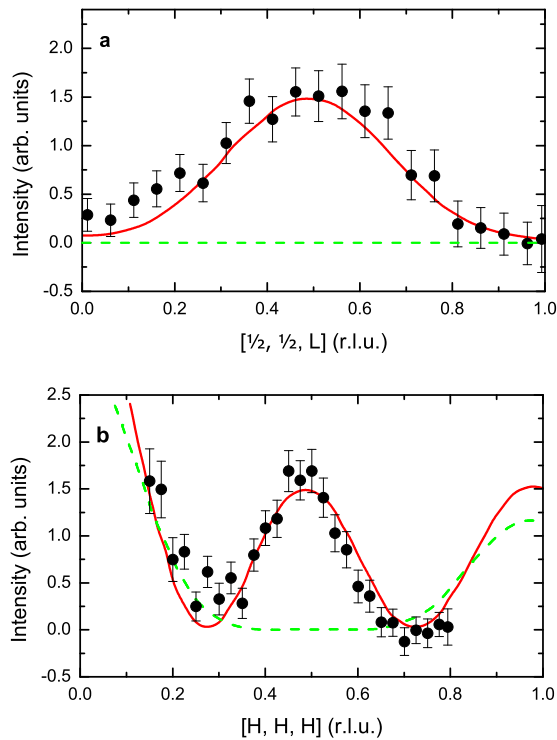}
\caption {(Color online)   The scan profiles along the [$\frac{1}{2}$, $\frac{1}{2}$, L] (a) and [H, H, H] (b) directions obtained from La$_{0.99}$Sr$_{0.01}$CoO$_3$ at $T=2$~K. Solid and broken lines are calculated intensities for the case of 7-site and 8-site magnetic clusters, respectively.  The typical cut width is 0.1 r.l.u. \label{INS_cut}}
\end{center}
\end{figure}

Eq.~\ref{single} gives rise to well defined intensity minima and maxima for different  scattering vectors $\mathbf{Q}$ (see Fig.~\ref{INS_map}~a), which are characteristic of the geometry of the magnetic cluster, contrary
to Eq.~\ref{average} where this information is smeared out due to the powder average in $Q$ space.
For instance, Eq.~\ref{single} predicts intensity
at $\mathbf{Q}=(\frac{1}{2},\frac{1}{2},\frac{1}{2})$ for a heptamer, in strong contrast to the octamer where zero intensity is expected at the same
$\mathbf{Q}$ position.
In order to verify the model, we mapped out the distribution of the intensity of the magnetic excitation at $\delta E=0.75$~meV within the (H,H,L) plane at $T=1.5$~K.
We found that the measured integrated intensity of the magnetic inelastic peak shows clear oscillatory behavior with a maximum intensity
at $\mathbf{Q}=(\frac{1}{2},\frac{1}{2},\frac{1}{2})$
(see Fig.~\ref{INS_map}~b) in full agreement with the calculated intensity for Co heptamers, Fig.~\ref{INS_map}~a.
This is also exemplified in Figs.~\ref{INS_cut}~a,b for measurements with the scattering vector $\mathbf{Q}$ along principal symmetry directions.
In all cases the fit to the heptamer configuration is rather good, whereas the octamer magnetic cluster model cannot satisfactorily fit the data along
the $\mathbf{Q}$= [H,H,H]  and  [$\frac{1}{2}$,$\frac{1}{2}$,L] directions.

Similar results were also obtained for other single crystal samples in this study
with ${x}\leqslant{0.01}$, $x=0.002$ and $x=0.005$.
In all cases at low temperatures we observe intense resolution limited inelastic peaks at $\delta E=0.75$~meV.
An additional INS peak at $\sim 0.6$~meV, due to the undoped \lco\ matrix, appears in the measurements at elevated temperatures, $T>30$~K.
The positions of the peaks do not depend on doping.
The excitations are dispersionless, indicating that intercluster interactions are weak and can be neglected.
Therefore, we conclude that in the light hole doping regime,
${x}\leqslant{0.01}$, a magnetoelectronic phase separation in the form of weakly interacting 7-site spin-state polarons in the nonmagnetic matrix is realized.

\subsubsection{\xlsco, $x\geqslant0.05$; decay of the spin-state polarons }

\begin{figure}[tb!]
\begin{center}
\includegraphics[width=0.95 \columnwidth]{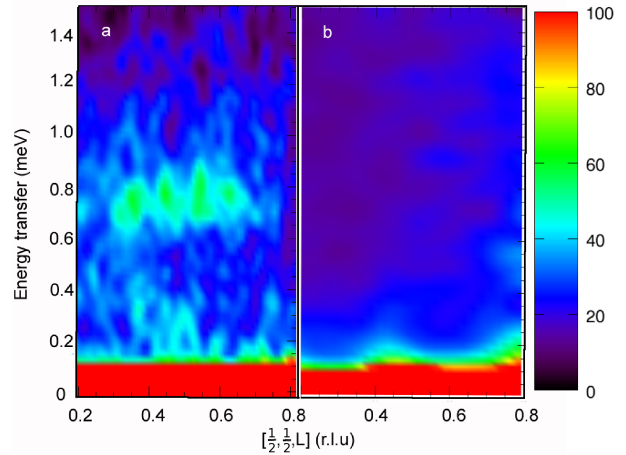}
\caption {(Color online) a) Two-dimensional map for La$_{0.99}$Sr$_{0.01}$CoO$_{3}$  plotted in energy-momentum space showing the inelastic peak at energy transfer $\delta E=0.75$~meV at $T=1.5$~K. b) The peak intensity collapses for the sample with higher hole concentration La$_{0.9}$Sr$_{0.1}$CoO$_{3}$. The signal is a section along $\mathbf{Q}$=($\frac{1}{2}$,$\frac{1}{2}$,L). The intensities are in arbitrary units.
\label{EvsQ}}
\end{center}
\end{figure}

The situation becomes more complicated if interpolaron interactions or/and interactions between individual magnetic cobalt ions exist (either IS Co$^{3+}$ or LS Co$^{4+}$) which belong to neighboring polarons.
In order to explore the behavior of the magnetic excitations at doping level above $x=0.04$ suggested as a lower limit of MEPS,\cite{He2} we measured INS spectra of two samples $x=0.05$ and $x=0.1$.
Although the quantitative comparison of INS peak intensities from the data obtained at different spectrometers is difficult, the qualitative tendency is obvious.
We found that the peak at $\delta{E}=0.75$~meV, which was the main magnetic feature of the system with ${x}\leqslant{0.01}$, is considerably suppressed at $x=0.05$ and totally disappears at $x=0.1$.
Our INS measurements provided no evidence for any dispersion of the 0.75~meV excitation, thus ruling out intercluster interactions as the origin of cluster-glass state.
Fig.~\ref{EvsQ} shows the observed inelastic intensity
around $\mathbf{Q}=(\frac{1}{2},\frac{1}{2},\frac{1}{2})$ obtained
from \xlsco, $x=0.01$ and 0.1.
The comparison suggests that the isolated spin-state polarons decay rapidly with hole-doping and can be hardly detected at ${x}\geqslant{0.05}$.

Note, that the high temperature INS peak at $E=0.6$~meV remains constant at increasing $x$ and then vanishes completely for ${x}\sim{0.1}$.
Here we would like to refine the corresponding part of \xlsco\ phase diagram.
The recently proposed magnetic phase diagram\cite{Wu1,He}
includes one more region in addition to the SG and CG-FM states.\cite{Itoh94}
This is the spin-state transition (SST) region at the very left side, ${0}\leqslant{x}\lesssim{0.01}$, of the diagram.
$T_{\textsc{sst}}$ is proposed to rapidly fall down from $T_{\textsc{sst}}(x=0)\thickapprox100$~K to $T_{\textsc{sst}}(x\thicksim 0.01) = 0$.
This is in contrast to the phase diagram proposed in Ref.~\onlinecite{G2}, where $T_{\textsc{sst}}\thickapprox 100$~K remains roughly constant for $0<x<0.15$.
The $T_{\textsc{sst}}$ of the thermally induced spin-state transition (or rather crossover) is determined by energy gap from the LS ground state to a first excited magnetic state of Co$^{3+}$ ions in a \textit{nondisturbed} \lco\ matrix.
The energy gap in undoped parent compound \lco\ was determined by means of inelastic neutron scattering and turns out to be $\sim 10.3$~meV.\cite{Podlesnyak1}
It follows from our current INS measurements that this LS $\rightarrow$ HS energy gap and hence $T_{\textsc{sst}}$ does not depend on Sr-doping.
This is indeed quite natural: it is not plausible that the doping on the level of several spins per thousand nonmagnetic ions would collapse the magnetic state of an entire system.
Therefore, we conclude that $T_{\textsc{sst}}\thickapprox100$~K in the main part of this region, and rather rapidly vanishes at $x \sim 0.1$, when polarons start to strongly overlap.

\begin{figure}[tb!]
\begin{center}
\includegraphics[width=0.95 \columnwidth]{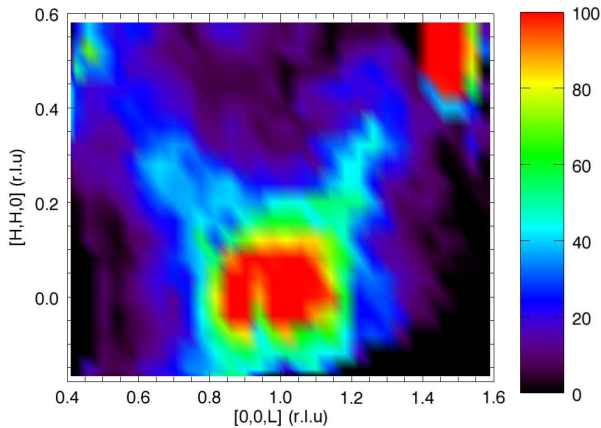}
\caption {(Color online) The magnetic elastic diffuse scattering $I_m=I(1.5\textrm{K})-I(100\textrm{K})$ measured on IRIS around FM  $\mathbf{Q}=$~(0,0,1) for La$_{0.9}$Sr$_{0.1}$CoO$_{3}$ showing anisotropic intensities along [H,H,H] directions.  The intensities are in arbitrary units.
\label{Elastic}}
\end{center}
\end{figure}

The data obtained from \xlsco, $x=0.1$ also reveal the clear presence of elastic diffuse scattering around the ferromagnetic $\mathbf{Q}=$~(0,0,1) wave vector, which was not observed at $x=0.01$.
To confirm the magnetic origin of the diffuse intensities we mapped out the elastic scattering at two temperatures 1.5 and 100~K.
Fig.~\ref{Elastic} shows a difference in the intensities obtained at temperatures 1.5 and 100~K, $I_m=I(1.5\textrm{K})-I(100\textrm{K})$.
The elastic scattering, indicative of static FM correlations, has a highly anisotropic shape.
The diffuse intensity extends along the (1,1,1) direction all the way
to ($\pm\frac{1}{2}$,$\pm\frac{1}{2}$,$1\pm\frac{1}{2}$,).
Note, that similar elastic diffuse scattering was reported by
Phelan \etal~\cite{Phelan,Phelan2}
Their spin-polarized measurements also proved that the observed intensities are dominantly magnetic in nature.\cite{Phelan2}
Moreover, they reveal static incommensurate magnetic correlations not observed in our measurements.
These results are indicative of coexisting and competing FM and AFM correlations.
For the hole-doping concentration ${x}\geqslant{0.1}$ the FM interactions become static, suggesting that relatively large FM clusters are formed at the expense of spin-state polarons.

\subsection{Magnetization}

Substitution of La$^{3+}$ with Sr$^{2+}$ provides hole doping and creates a mixed Co$^{3+}$-Co$^{4+}$ system.
However, the Sr$^{2+}$ ion has a bigger ionic radius
(1.18~\AA~ and 1.032~\AA, for Sr$^{2+}$ and La$^{3+}$, respectively), that can also locally distort the crystal structure.
Therefore, the Sr$^{2+}$  substitution leads not only to the hole injection,
 but may also change $\Delta_{cf}$ in doped clusters.
In order to understand the role of local structure distortions in the observed magnetic effects we measured the dc magnetic susceptibility of LaCoO$_3$ doped with Sr$^{2+}$, Ca$^{2+}$ and Y$^{3+}$.

\begin{figure}[tb!]
\begin{center}
\includegraphics[width=0.95 \columnwidth]{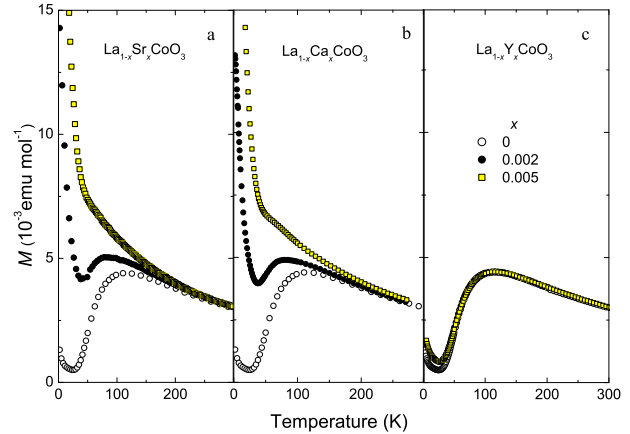}
\caption {(Color online) Temperature dependence of the dc magnetic susceptibility for La$_{1-x}M_{x}$CoO$_{3}$, $M=$~Sr, Ca, Y taken at $H=1$~T.  \label{YCa_doped}}
\end{center}
\end{figure}

Yttrium is an isovalent ion to lanthanum, however, with a much smaller ionic radius (0.90~\AA).
One can expect that distortions due to Y$^{3+}$ doping would be even larger compared to strontium substitution.
On the other hand, Ca$^{2+}$ has an ionic radius of 1.00~\AA, close
to La$^{3+}$.
The magnetic susceptibility curves show a pronounced doping effect in case of Sr and Ca doping (Fig.~\ref{YCa_doped}~a,b).
The susceptibility exhibits a strong increase at low temperatures compare to parent compound \lco.
At increasing temperature the susceptibility goes through the broad maximum at $T\sim{100}$~K indicating thermal activation of Co$^{3+}$ HS state ions.
On the other hand, in case of Y$^{3+}$ doping very little changes of the magnetic susceptibility were observed compared to the
undoped \lco\ (Fig.~\ref{YCa_doped}~c).
This is unambiguous evidence that the substitution of La$^{3+}$ for Sr$^{2+}$ provides mainly holes to the system without creation of a sizable crystal field distortion in the doped clusters.
The hole doping is the main origin for the observed low temperature magnetic anomalies.
This is unambiguous evidence that the substitution of La$^{3+}$ by Sr$^{2+}$ acts  mainly by providing holes to the system, and not by  a  crystal field distortion in the doped clusters.
Thus the hole doping is the main origin for the observed low temperature magnetic anomalies.

\begin{figure}[tb!]
\begin{center}
\includegraphics[width=0.95 \columnwidth]{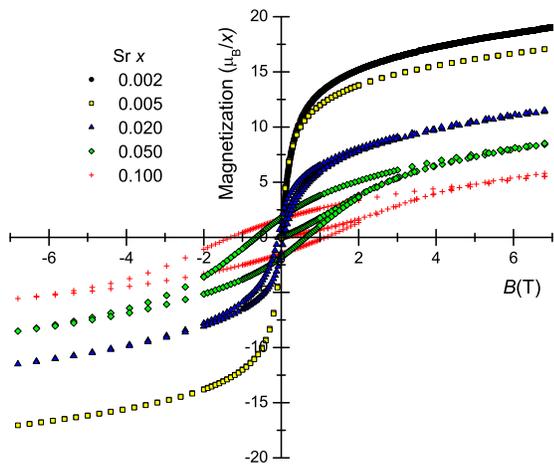}
\caption {(Color online) Field dependencies of the magnetization per dopant $x$ for \xlsco\ measured  on powder samples at $T=2$~K.
\label{Magnetization}}
\end{center}
\end{figure}

\begin{figure}[tb!]
\begin{center}
\includegraphics[width=0.95 \columnwidth]{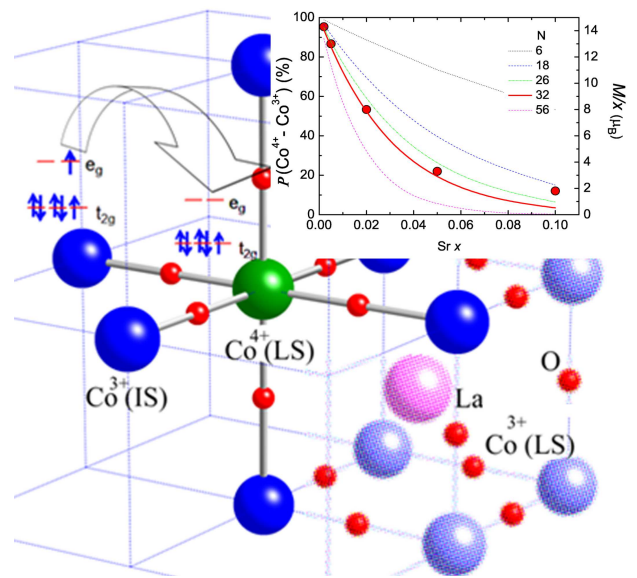}
\caption {(Color online) Schematic view of the spin-state polaron. Light blue spheres denote Co$^{3+}$ in LS state matrix (only one unit cell is shown for clarity) while dark blue ones are switched to the IS state via nearest neighboring Co$^{3+}$ - Co$^{4+}$ interaction (see text).
The curves in the inset show  the probability $\mathcal{P}(x)$ for the given Co$^{4+}$ ion to find $\mathcal{N}$ nearest neighbors (in one; two; three; four; and five shells) in  Co$^{3+}$ state (left axis). The circles show the concentration evolution of the saturated magnetic moment per dopant as obtained from magnetization measurements (right axis). \label{Polaron}}
\end{center}
\end{figure}

The low temperature field dependence of the magnetization per doped hole, (or, in other words, per Co$^{4+}$) for different doping concentrations is shown in Fig.~\ref{Magnetization}.
It is worth to mention that Co$^{4+}$ is expected to be in LS state
($t_{2g}^5e_g^0$), thus the expected magnetic moment
is $M/x=1$~$\mu_{\textsc{b}}$.
In order to estimate an effective magnetic moment we fitted measured magnetization $M(H)$ with a combination of the conventional Brillouin function B$_S(y)$ and a field-linear term,
$M(H)={x}\mu_{\textsc{b}}\cdot{gS}\cdot{B_S(y)}+\chi_{0}H$,
$y=(g\mu_{\textsc{b}}SH)/(k_{\textsc{b}}T)$.\cite{Yam1}
The resulting values $M/x$ are shown as full dots in the inset of Fig.~\ref{Polaron}.
For the lowest doping $x=0.002$ the magnetization curves correspond to the saturation moment $M/x\sim{13}-15$~$\mu_{\textsc{b}}$/hole, which is much higher than one can expect from single Co$^{4+}$ or Co$^{3+}$ in any spin state.
This result combined with the INS data for the lightly doped cobaltites
${x}\leqslant{0.01}$ fully supports our spin-state polaron model.
Each injected hole triggers off neighboring Co$^{3+}$ to the IS magnetic state creating a magnetic cluster with $M/x=13$~$\mu_{\textsc{b}}$.
A reasonable mechanism for such a resonant state with a hole "dressed'' by the magnetic cloud was proposed by Louca and Sarrao.\cite{Louca2}
Fig.~\ref{Polaron} represents a schematic view of such a spin-state polaron.
Neighboring LS Co$^{4+}$ and IS Co$^{3+}$ ions share an $e_g$ electron by swapping configuration that would be energetically favorable for $e_g$ hopping.
The $t_{2g}$ electrons, in turn, couple ferromagnetically via double exchange interaction thus forming a giant magnetic moment.

Surprisingly, the magnetization curves tend to saturate at lower values $M/x$ upon hole doping indicating a rapid reduction of the magnetic moment per hole. For low $x$ (well below the percolation threshold) it is natural to expect that the phase-separated system would consist of almost noninteracting spin-state polarons with giant magnetic moment, separated by the nonmagnetic \lco\ matrix.
The value of the moment $M/x$ should remain roughly constant since the number of polarons is proportional to $x$ till such polarons start to overlap. However, as one can see in  Fig.~\ref{Magnetization} and in inset of Fig.~\ref{Polaron}, $M/x$ rapidly drops down to $\sim 2-3$~$\mu_{\textsc{b}}$/hole,  the values which are characteristic of the magnetic moment of a single cobalt ion.
This result provides further evidence that the spin-state polarons collapse as $x$ is increased.

\begin{figure}[tb!]
\begin{center}
\includegraphics[width=0.95 \columnwidth]{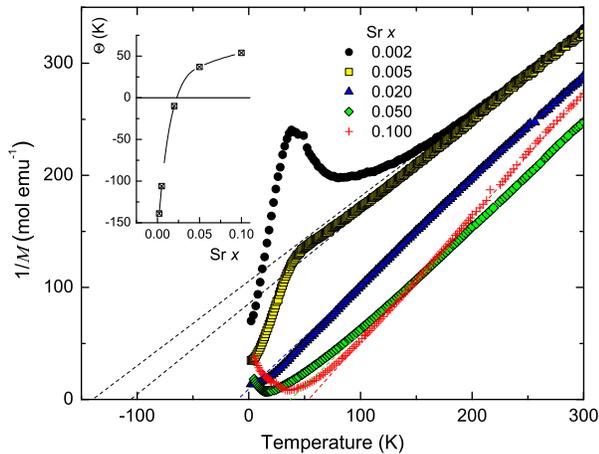}
\caption {(Color online) Inverse dc magnetic susceptibility versus temperature for \xlsco. Broken lines are Curie-Weiss linear fits to the high temperature data. The inset shows the concentration evolution of the paramagnetic Curie temperature $\Theta$. \label{Inverse_Xi}}
\end{center}
\end{figure}

The paramagnetic Curie temperature $\Theta$ is another value to elucidate interspin interactions.
The value of $\Theta$ is an arithmetic average of the interspin coupling constants $J_{RR{^\prime}}$,
\begin{equation}
\Theta=\frac{S(S+1)}{3k_{\textsc{b}}} \frac{1}{N} \sum_{RR{^\prime}} J_{RR{^\prime}} , \label{Theta}
\end{equation}
where the sum is over all $N$ interacting spins  (see, for instance,
Ref.~\onlinecite{Czachor}).
That is, in case of several subsystems with competing magnetic interactions in a phase separated compound the $\Theta$ value provides an indicator of their relative strength.
The temperature dependence of the inverse magnetic susceptibility was measured in field $H=1$~T on heating from low temperatures after zero field cooling (Fig.~\ref{Inverse_Xi}).
Although the spin-state crossover makes a fit to the Curie law difficult at low temperatures, we were able to estimate the paramagnetic Curie temperature
$\Theta$ by fitting the range $T>150$~K.
As shown in the inset of the Fig.~\ref{Inverse_Xi}, $\Theta$ quickly increases from negative to positive values with increasing $x$, crossing zero at about $x\sim 0.03-0.04$.
This suggests that the strength of competing FM-AFM contributions to $\Theta$ strongly depends on doping level and FM-AFM correlations become comparable around the MEPS low limit, $x\sim 0.04$.

As we already mentioned in the introduction, three different types of magnetic interactions compete in lightly hole-doped \xlsco: 1) intrapolaron interactions between IS Co$^{3+}$ and HS Co$^{4+}$; 2) interpolaron interactions as well as polaron - undoped matrix interactions at elevated temperatures; and 3) interactions between individual Co spins from different spin-state polarons (for low $x$) or magnetic clusters (for higher $x$).
Intrapolaron interactions are ferromagnetic via double exchange interaction and cause of the giant moment.
The negative values of $\Theta$ in the lightly doped system ${x}\lesssim{0.04}$ as well as neutron scattering data\cite{Phelan1,Phelan2} indicate that thermally-excited magnetic (HS/IS) states interactions are mainly AFM.
The hysteresis loops which are observed in \xlsco\ for ${x}\geqslant{0.02}$ emphasize that the FM interactions become prominent starting from these elevated $x$.
Note, that coexisting and competing FM and AFM correlations in hole-doped cobaltites were proved by various different techniques, such as neutron scattering,\cite{Phelan1,Phelan2} specific heat,\cite{He2,He3} magnetic susceptibility,\cite{Andro} NMR,\cite{Kuhns} and $\mu$SR.\cite{Giblin1}

It is apparent that the interactions between individual cobalt ions from neighboring polarons is negligible in the system with low $x$.
They become more and more important with increased doping.
As a consequence of the statistical clustering of Co$^{4+}$, the number of isolated polarons rapidly decreases in favor of interacting ones.
To estimate how such Co-Co interactions are related to the magnetic properties of \xlsco, we apply a simple geometrical consideration.
Since the spin-state polaron has octahedral shape (see Fig.~\ref{Polaron}), it can be considered as isolated when for the given central Co$^{4+}$ ion all $\mathcal{N}$ neighbors from the first to the fourth shell are in the Co$^{3+}$ state.
Otherwise, two polarons share a common IS Co$^{3+}$.
This gives us six nearest neighbors along the edge, plus twelve ions along the face diagonal, plus eight along the body diagonal and six more along the edge at two unit cells distance - altogether 32 "blocked'' sites.
For any given Co$^{4+}$ ion the probability $\mathcal{P}(x)$ to find all  $\mathcal{N}$ neighbors in Co$^{3+}$ is $\mathcal{P}$(Co$^{4+}$-Co$^{3+})=(1-x)^{\mathcal{N}}$.
As one can see from the inset in Fig.~\ref{Polaron}  the value of the magnetic moment as a function of $x$ is indeed best scaled with the curve which takes into account all neighbors from the first to the fourth shell.
This suggests that the FM interactions between neighboring IS Co$^{3+}$-Co$^{3+}$ out of different polarons overcome the AFM correlations of polarons themselves when the interpolaron distance is reduced to the order of two unit cells.
Since the Co$^{3+}$ ions that are situated between two Co$^{4+}$ polaron centers experience both FM intra- and AFM interpolaron interactions they turn out to be frustrated.
The decay of polarons on further increasing ${x}\geqslant{0.05}$ gives rise to the larger-scale clusters with competitive FM-AFM interactions (i.e. cluster-glass state), which is confirmed by the decrease of the magnetic moment with simultaneous increasing of hysteresis loop.
Note, that the clear deviation of the $M/x$ from the calculated curve above the critical  doping $x \gtrsim 0.05$ can be also explained by the growing magnetic contribution from  the larger-scale clusters.

Our findings are in good agreement with conclusions of He \etal~\cite{He3,He2}.
They argued that  the phase separation i) is restricted to a well-defined doping range, $0.04<x<0.22$, and ii) is driven solely by inevitable
local compositional randomness at nanoscopic length scale.
Combining these ideas with the present neutron scattering and magnetization data, we can suggest that the hole concentration $x\sim0.04$ is a crossover from polaron type of magneto-electronic inhomogeneity to a state with ferromagnetic spin clusters which are formed at the expense of spin-state polarons.

\section{CONCLUSIONS}

We have performed a comprehensive study of inelastic neutron scattering and magnetization in \xlsco\ single crystals, $0<x<0.1$.
We conclude that the magnetoelectric phase separation for the lightly hole-doped cobaltites, $x<0.04$, has the form of the seven-site octahedral spin-state polaron and thus is an electronically driven process as opposed to the doping-driven phase separation at $x>0.04$.
We confirm that FM-AFM frustrated interactions coexist over wide composition range.
The agreement between experiment and our simple statistical calculations implies that with increasing $x$ the strong ferromagnetic correlations are associated with Co-Co rather than interpolaron interactions.
According to our inelastic neutron scattering measurements the low spin -- high spin energy gap of Co$^{3+}$ matrix, and hence $T_{\textsc{sst}}$, remains roughly constant with Sr-doping up to $x\sim0.05$.

\appendix*
\section{}
Considering only nearest-neighbor coupling $\mathcal{J}$, the Heisenberg exchange Hamiltonian is given by $\mathcal{H}_{ex}=-2 \mathcal{J} \textbf{S}_1 \cdot \textbf{S}_{\textsc{a}}$, where $\textbf{S}_1=1/2$ and $\textbf{S}_{\textsc{a}}=S_2+S_3+ \ldots +S_7$. The total spin is $\textbf{S}=\textbf{S}_1 + \textbf{S}_{\textsc{a}}$. The heptamer states are therefore defined by the wave functions $|S_1,S_{\textsc{a}},S\rangle$ with $0 \leqslant S_{\textsc{a}} \leqslant 6$ and $|S_{\textsc{a}} - 1/2| \leqslant S \leqslant (S_{\textsc{a}} + 1/2)$. With this choice of spin quantum numbers, the Hamiltonian is diagonal; thus, the energy eigenvalues can readily be derived, $E(S_{\textsc{a}},S)=-\mathcal{J}[S(S+1)- S_{\textsc{a}}(S_{\textsc{a}}+1)-3/4]$.
The Co-Co coupling $\mathcal{J}$ is ferromagnetic via the double exchange mechanism.\cite{Louca2} The ground state of the Co heptamer is therefore the state with maximum spin quantum numbers, namely $|S_1,S_{\textsc{a}},S \rangle = |1/2,6,13/2\rangle$ with energy $E(1/2,6,13/2)=-6\mathcal{J}$. The first-excited state is then $|1/2,5,11/2\rangle$ with $E(1/2,5,11/2)=-5\mathcal{J}$, i.e., it is separated from the ground state by the energy $\mathcal{J}$.
The exchange coupling $\mathcal{J}$ of Co$^{3+}$ oxides is of the order of 20~meV, thus, the first-excited heptamer state lies far above the energy window covered by the present experiments. \lco\ crystallizes in the rhombohedral space group $R\overline{3}c$.
Therefore, the ground state is split by the trigonal ligand field  into seven doublets $|\pm M\rangle$ $(0 \leqslant M \leqslant S)$.
 Thus we identify the peak observed at 0.75~meV with the lowest transition $|\pm13/2\rangle \rightarrow |\pm 11/2 \rangle$. In fact, the observed temperature dependence of its intensity supports this interpretation.\cite{Podlesnyak2}

\begin{acknowledgments}
The authors thank N. Baranov for fruitful discussions. This work is partly based on experiments performed at the Swiss spallation neutron source SINQ,  Paul Scherrer Institute,  Switzerland and ISIS facility, UK. This research at Oak Ridge National Laboratory's Spallation Neutron Source was sponsored by the Scientific User Facilities Division, Office of Basic Energy Sciences, U. S. Department of Energy. Research at ORNL is partly sponsored by the Materials Sciences and Engineering Division, Office of Basic Energy Sciences, US Department of Energy.  ORNL is managed by UT-Battelle, LLC, under contract DE-AC05-00OR22725 for the U.S. Department of Energy. The authors are grateful for the local support staff at SNS, ISIS and PSI. The work of E.P. was partly supported by the NCCR program MaNEP.
\end{acknowledgments}

\bibliographystyle{apsrev}
\bibliography{hole_doping}

\begin{thebibliography}{41}
\bibitem[{eae({\natexlab{a}})}]{eae1}
\bibinfo{howpublished}{Currently, there are hot debates concerning the magnetic
  properties of this {"}nonmagnetic{"} low-temperature state (see, for
  instance, Ref.~\onlinecite{Yam1}).}
\bibitem[{eae({\natexlab{b}})}]{eae}
\bibinfo{howpublished}{The value of 10~meV was obtained from INS measurements
  and includes lattice relaxations \cite{Kyomen}. Without these relaxations an
  energy difference of at least 50~meV between the LS and the HS was obtained
  from the cluster calculations \cite{Haverkort1}. Also note that the nature of
  the first excited state is still a matter of controversy which is out of the
  scope of this paper though (for details, see Ref.~\onlinecite{Podlesnyak1}
  and references herein).}
\end{thebibliography}

\end{document}